\documentstyle[fleqn]{article}
\newcommand{\eq}{\begin{equation}}
 \newcommand{\en}{\end{equation}}
 \newcommand{\eqn}{\begin{eqnarray}}
 \newcommand{\enn}{\end{eqnarray}}
 \newcommand{\nn}{\nonumber }
\newcommand{\g}{\gamma}
\def\g{\gamma}
\def\d{\delta}
\def\pa{\partial}

\def\ref#1{$\sp{#1)}$}

\def\pl#1#2#3{Phys.~Lett.~{\bf {#1}B} (19{#2}) #3}
\def\np#1#2#3{Nucl.~Phys.~{\bf B{#1}} (19{#2}) #3}

\begin{document}
\begin{titlepage}
\begin{flushright}
PSU-TH-180 \\
December 1996 
\end{flushright}
\begin{center}

{\bf SEVEN SPHERE AND THE EXCEPTIONAL  NONLINEAR SUPERCONFORMAL 
ALGEBRAS } \\
\vspace{1cm}
{\bf Murat G\"{u}naydin} \footnote{Work supported in part by the National Science Foundation under research   Grant No. PHY-9631332. 
\newline email: murat@phys.psu.edu}  \\
Physics Department, 104 Davey Lab. \\
Penn State University  \\   
 University Park, PA 16802, U.S.A. \\
\vspace{1cm}
  Talk given at the 30th International Symposium \\
on the Theory of Elementary Particles, \\
Buckow, Germany (August 27-31, 1996).     \\  
\vspace{1cm}
{\bf abstract} \\
\end{center}
\vspace{0.5cm}
The realizations of the exceptional non-linear (quadratically generated, 
or $W$-type) $N=8$ and $N=7$ superconformal algebras with $Spin(7)$ and $G_2$
affine  symmetry currents are reviewed. Both the $N=8$ and $N=7$ algebras 
admit unitary realizations in terms of a single boson and 
free fermions in ${\underline{8}}$ of $Spin(7)$ and ${\underline{7}}$ of $G_2$,
with the central charges $c_8=26/5$ and $c_7=5$, respectively. They can also be
realized over  the coset spaces $SO(8)\times U(1)/SO(7)$ and $SO(7)\times U(1)/G_2 $
 for some fixed  values of their
central charges, respectively. The coset space $SO(8)/SO(7)$ is  the seven-sphere $S^7$, 
whereas the space $SO(7)/G_2$ represents the seven-sphere with torsion. We conclude
with a discussion of
 a novel 'hybrid' method developed recently that yields
unitary realizations of the exceptional $N=8$ and $N=7$ algebras for all allowed
values of their central charges.

\end{titlepage}

\renewcommand{\theequation}{\arabic{section} - \arabic{equation}}
\section{Introduction}
\setcounter{equation}{0}

The $N$-extended quadratically non-linear superconformal algebras of the type introduced by Bershadsky 
and Knizhnik \cite{kn,be} involve generators of conformal dimension $2$, $3/2$
 and $1$ only. The anticommutators of  the supersymmetry generators of these algebras close into
the Virasoro generators, generators of the symmetry currents of dimension one and their bilinears.
 A complete classification of complex forms of 
simple reductive non-linear superconformal algebras was 
given in refs.~\cite{fl,bo}. Reductivity  means that these algebras linearise in 
the limit of infinite central charge. In this limit, the 
infinite-dimensional vacuum-preserving algebra becomes a finite superalgebra
containing the finite (global) conformal algebra and the  classification
 of such non-linear superconformal algebras \cite{fl,bo} reduces to the 
classification of finite-dimensional (global) superconformal algebras given in
ref.~\cite{gst}. There are three infinite classical families (for either the 
right- or the left-moving modes) , 
$$osp(N|2;{\bf R})~,\quad su(1,1|N)~,\quad osp(4^*|2N) $$
,a one-parameter family of  $N=4$ algebras, and two exceptional 
superconformal algebras with $N=7$ and $N=8$ supersymmetries. 
 The vacuum-preserving subalgebras of 
the $N=7$ and $N=8$ exceptional superconformal algebras in the limit of 
infinite central charge are the exceptional finite Lie superalgebras $G(3)$ 
and $F(4)$ (in  Ka\v{c}'s notation \cite{kac}), respectively. \footnote{See  
refs.~\cite{snr,dvn} for details about the $G(3)$ and $F(4)$. Another real 
form of $F(4)$ corresponds to $N=2$ superconformal symmetry in five space-time 
dimensions \cite{mg1}.} In this talk I will review the realizations of the exceptional
nonlinear superconformal algebras. Following my joint work with Sergei 
Ketov \cite{gk} I will first give their realizations  in terms of a single boson
and $N$ fermions and  then study their realizations over the coset spaces
$SO(8)\times U(1)/SO(7)$ and $SO(7)\times U(1)/G_2\,$, respectively. Both 
of these realizations give some definite values for the central charges of
the respective algebras. I will conclude with the discussion of their realization
via a "hybrid" method given recently in my joint work with Behzad Bina \cite{bg}. This hybrid
method involves fermions in the coset of some supergroup, a scalar field (dilaton) and dimension
one currents corresponding to the symmetries of these algebras and yields realizations
for all allowed values of the central  charge determined by the level of the
symmetry current algebra.  

For discussion of physical motivations , possible applications to string theory as well
as further details of the results reported below and further references we refer to references \cite{gk} and \cite{bg}.

\section{Exceptional $N=7$ and $N=8$ Superconformal Algebras}
\setcounter{equation}{0}

It is most convenient to represent the commutation and anticommutation relations \cite{bo}
of the generators of the $N=8$ and $N=7$
non-linear superconformal algebras  as operator products of chiral fields ( either in the left or the
right moving sector)  defined in terms of these  generators \cite{gk}. These algebras 
can be obtained either via a Drinfeld-Sokolov-type reduction from affine 
versions of the exceptional Lie superalgebras $F(4)$ and $G(3)$, respectively
\cite{imp}, or by purely algebraic methods~\cite{fl,bo}. The $N=8$ algebra 
contains eight supercurrents $S^M$ of conformal dimension $3/2$, and $21$ 
symmetry currents of $Spin(7)$ under which the supercurrents transform in the 
{\it spinor} representation. The $N=7$ algebra has $7$ supercurrents and $14$
 symmetry currents of $G_2$. Because of  their 
non-linearity, the `vacuum-preserving' algebra, generated by the modes 
$S^M_{\pm 1/2}$  is {\it not} finite.
However in the limit of infinite central charge it becomes finite.

\subsection{Exceptional $N=8$  superconformal algebra}

The bosonic part of the $N=8$ algebra is a semi-direct sum of the affine
algebra $\widehat{so(7)}_k$ of level $k$ and the Virasoro algebra. Their 
OPEs are given as
\eqn
 T(z) T(w) &\sim &\frac{c/2}{(z-w)^4} + \frac{2T(w)}{(z-w)^2}  
+\frac{\partial T(w)}{z-w} \\ \nn
 T(z)T^{mn}(w)&\sim &\frac{T^{mn}(w)}{(z-w)^2} + \frac{\partial T^{mn}(w)}{z-w}  \\
T^{mn}(z)T^{pq}(w) &\sim &\frac{-i}{z-w} \{ \delta^{np}T^{mq}(w)  
+ \delta^{mq}T^{np}(w)  
- \delta^{mp}T^{nq}(w) 
\\ \nn 
&& -  \delta^{nq}T^{mp}(w) \}
 +\frac{k}{(z-w)^2}\{ \delta^{mp}\delta^{nq} -  \delta^{mq}\delta^{np}\} 
 \nn
\enn
where the adjoint representation of $SO(7)$ is labeled by a pair of antisymmetric indices, 
$m,n,\ldots=1,2,\ldots,7$. 
Since the supercurrents $S^M(z) $ transform in the spinor representation 
of $Spin(7)$ and have spin 3/2, we have 

\eqn
T(z) S^M(w) \sim  \frac{\frac{3}{2}S^M(w)}{(z-w)^2}+\frac{\pa S^M(w)}{z-w} \\ \nn
T^{mn}(z)S^M(w) \sim \frac{-i}{2} \frac{\g^{mn}_{MN}}{(z-w)} S^N(w)
\enn

where $\g^{mn}=\frac{1}{2} [\g^m,\g^n ]$ with $\g^m$  the $8\times 8$ gamma matrices in seven dimensions. These gamma matrices
 can be written in terms of the octonionic structure
constants \cite{gg,cdfn,wn,dgt,gk}.\footnote{There
exists two different realizations of the gamma matrices in seven dimensions in terms of the
octonionic structure constants. We shall denote them as $\g^m$ and $\tilde{\g}^m $  \cite{gk}.
The antisymmetric products of gamma matrices are defined  with
unit weight, {\it viz.}
$ \g^{ij\cdots k}=\g^{[i}\g^{j}\cdots \g^{k]}$.} 
The  non-trivial OPE's are the ones corresponding to the products of
 $N=8$ supersymmetry generators which read as follows:

\eqn
S^M(z)S^N(w) &\sim &\frac{8k(k+2)}{3(k+4)}\frac{\delta^{MN}}{(z-w)^3} 
+\frac{2T(w)}{z-w}\delta^{MN} \\ \nn
&& - \frac{\delta^{MN}}{3(k+4)}\frac{T^{mn}T^{mn}(w)}{z-w}
- \frac{1}{12(k+4)}\g^{MN}_{mnpq}
\frac{:T^{mn}T^{pq}:(w)}{z-w} \\
&& + \frac{2i(k+2)}{3(k+4)}\g^{MN}_{mn}\{ \frac{T^{mn}(w)}{(z-w)^2} 
 + \frac{\pa T^{mn}(w)}{2(z-w)}\} \nn
\enn
where $M,N = 1,\ldots,8$, and $m,n\ldots=1,\ldots,7$. 

The Jacobi `identities' require that 
the central charge $c$ of the $N=8$ algebra be  related to the level $k$ 
of affine $Spin(7)$ as
follows: \cite{fl,bo,gk}.

\eq 
c= c_8\equiv 4k + \frac{6k}{k+4} \equiv \frac{2k(2k+11)}{k+4}~,
\en

\subsection{Exceptional $N=7$ superconformal algebra}

The exceptional $N=7$ non-linear superconformal algebra is similar to the 
$N=8$ algebra having the gauge group $G_2$ instead of $Spin(7)$ and seven 
supercurrents. We shall denote the generators of $G_2$ as $G^A$ 
and  the seven-dimensional representation matrices of $G_2$ 
 as $M^A$ $(A=1,2,\ldots,14)$. The matrices $M^A$ satisfy the identities \cite{gg,bo,gk}

\eq
tr (M^AM^B) = 2 \d^{AB}  
\en
\[
M^A_{ij}\;M^A_{kl} =  \frac{2}{3}(\d_{il}\d_{jk}-\d_{ik}\d_{jl})
-\frac{1}{3}C_{ijkl}
\]   \nn

where $C_{ijkl}$ is the $G_2$ invariant completely antisymmetric tensor .
The seven supercurrents $S^i(z)$ transform in $7 $ of $G_2$, and 
satisfy the OPEs

\eqn
T(z) S^i(w) \sim \frac {\frac{3}{2}S^i(w)}{(z-w)^2} 
+\frac{\pa S^i(w)}{z-w}  \\ \nn
 G^A(z)S^i(w) \sim \frac{1}{z-w}M^A_{ij}\,S^j(w)  
\enn

The most important OPE's are again the ones between the  $N=7$ supersymmetry
generators

\eqn
S^i(z)S^j(w) &\sim &\frac{k(3k+5)}{k+3}\frac{\d^{ij}}{(z-w)^3} 
+\frac{3k+5}{k+3}M^A_{ij} \{\frac{G^A(w)}{(z-w)^2}+\frac{1}{2} \frac{\pa
G^A(w)}{z-w} \} \nn
\\ 
&&+\frac{\d^{ij}}{z-w} \{2T(w) - \frac{1}{k+3} G^AG^A(w) \}  
\\ 
&&+\frac{3}{4(k+3)} \frac{\{ M^AM^B +M^BM^A \}^{ij} G^AG^B(w)}{(z-w)}
 \nn
\enn
The central charge is given by \cite{fl,bo,gk}
$$ c = c_7 \equiv \frac{9}{2}k + \frac{2k}{k+3} \equiv 
\frac{k(9k+31)}{2(k+3)}~ $$

\section{Coset Space Realizations}
\setcounter{equation}{0}

The structure of the exceptional nonlinear $N=8$ and $N=7$ superconformal algebras is such that they
do not admit realizations over an infinite family of coset spaces  $G/H$ in contrast to superconformal
algebras with $N \leq 4$  \cite{gk}. Starting with a most general Ansatz for a coset space realization of 
these algebras one finds that such a realization is possible for a unique compact coset space for each of the exceptional algebras \cite{gk}  , namely $SO(8)\times U(1)/SO(7)$ for the $N=8$ algebra and 
$SO(7) \times U(1) / G_2$ for the $N=7$ algebra.

\subsection{A construction of the exceptional $N=8$ superconformal
algebra over the coset space  $SO(8)\times U(1)/SO(7)$}
\setcounter{equation}{0}

Consider  the affine algebra $\widehat{so(8)}_{\hat{k}} \oplus 
\widehat{u(1)}$, defined by the OPEs 

\eqn
\hat{J}^{ab}(z)\hat{J}^{cd}(w) & \sim & 
\frac{2}{z-w}\{ \d^{bc}\hat{J}^{ad}(w)
 + \d^{ad}J^{bc}(w)  
 - \d^{ac}\hat{J}^{bd}(w) -  \d^{bd}\hat{J}^{ac}(w) \}
 \nn \\
&& 
 -\frac{4\hat{k}}{(z-w)^2} \{ \d^{ac}\d^{bd}-\d^{ad}\d^{bc}
\} 
\enn
and

\eq
\hat{J}^8(z)\hat{J}^8(w) \sim \frac{\hat{k}_1/2}{(z-w)^2}
\en

where $a,b,\ldots=1,2,\ldots,8$ . The  $\hat{k}_1$ is a normalisation parameter
of the $U(1)$ current $\hat{J}^8$. 
The currents $\hat{J}^m=\pm i\hat{J}^{m8}$  ,
$m=1,\ldots,7$, belonging to the coset $SO(8)/SO(7)$ satisfy the OPE

\eq
\hat{J}^m(z)\hat{J}^n(w) = \frac{4\hat{k}}{(z-w)^2}\d^{mn} 
+\frac{2\hat{J}^{mn}(w)}{z-w}\, +\hat{J}^m\hat{J}^n(w)+\ldots
\en

Associated with the  currents $\hat{J}^a = (\hat{J}^m,\hat{J}^8)$ of the coset  $SO(8)\times U(1)/SO(7)$ 
we introduce free fermionic fields $\psi^a =(\psi^m , \psi^8) \, , a,b,\ldots =1,2,\ldots ,8$ 

\eq
\psi^a(z)\psi^b(w) = \frac{1/2 \delta^{ab}}{z-w} + (z-w)\pa\psi^a\psi^b(w) 
\en

One finds that a consistent solution to all the contraints exists and leads to the 
following expressions for the generators of the $N=8$ algebra in terms of 
the coset space currents and fermions \cite{gk}

\[
T^{mn} =  - \frac{i}{2} \{ \hat{J}^{mn} + \bar{\psi}
 \tilde{\gamma}^{mn} \psi \} 
\] \nn
\eq
S^m =  i\sqrt{\frac{7}{33}}\,\g^m_{ab}\psi^a\hat{J}^b,\qquad
 S^8 =  \sqrt{\frac{7}{33}}\, \psi^a\hat{J}^a  
\en
\eqn
T   &=&   \frac{1}{132} \{ 7 \hat{J}^a\hat{J}^a
- \hat{J}^{mn}\hat{J}^{mn} -42 \bar{\psi}\pa\psi 
 + \frac{3}{2}\hat{J}^{mn} (\bar{\psi}\tilde{\g}^{mn}\psi)
 - (\bar{\psi}\tilde{\g}^{mn}\psi)(\bar{\psi}\tilde{\g}^{mn}\psi)  \}
 \nn
\enn
with central charge $c=84/11$ corresponding to the level $k=\hat{k} +1=3/2$.

\subsection{A construction of the exceptional $N=7$ superconformal algebra
                  over the coset space $SO(7)\times U(1)/G_2$}

The automorphism group $G_2$ of octonions is a 
14-dimensional subgroup of $SO(7)$. Under $G_2$, the adjoint representation of
$Spin(7)$ decomposes as  $21= 14 +7 $. One can choose a basis for
 $G_2$ such that its generators $G^{ij}$ 
can be expressed in terms of the generators $J^{ij}$ of $SO(7)$ in a simple 
form \cite{wn,gn,gk}:

\eq
 G^{ij} = \frac{1}{2} J^{ij} + \frac{1}{8}C^{ij}_{~~kl} J^{kl}
\en

The $G^{ij}$ are not all linearly independent and satisfy the constraints

\eq
 C_{ijk}G^{jk}=0
\en
 
where $C_{ijk}$ are the completely antisymmetric structure constants of the octonions.
The remaining seven generators  of  $SO(7)$  can be chosen as

\eq
A^i=\frac{1}{2} C^{ijk} J^{jk}
\en

They are associated with the seven-dimensional coset space $SO(7)/G_2$. 
Hence we have the decomposition \cite{gn}

\eq
J^{ij}=\frac{4}{3}G^{ij} + \frac{1}{3}C^{ijk}A^k
\en

Note that the coset space $SO(7)/G_2$ is {\it not} a 
symmetric space. The symmetric space $SO(8)/SO(7)$  can be identified with the
 round seven-sphere  $S^7$  and the space $SO(7)/G_2$  as the   
seven-sphere with torsion.

For  the $N=7$ algebra  we  start 
with  the affine algebra $\widehat{SO(7)}_{\hat{k}}$ . Denoting the currents associated with 
the coset  space $SO(7)/G_2$ and the group
 $G_2$ as  $\hat{A}^m(z)$  and $\hat{G}^{mn}(z)$ ,respectively, we have

\eq
\hat{A}^m(z) = \frac{1}{2} C^{m}_{~np}\hat{J}^{np}(z) 
\en 
\[
\hat{G}^{mn}(z)\equiv \frac{1}{2} \hat{J}^{mn}(z) 
+ \frac{1}{8}C^{mn}{~~pq}\hat{J}^{pq}(z)
\] \nn

The coset space currents satisfy the  OPE

\eq
 \hat{A}^m(z) \hat{A}^n(w) = \frac{-12\hat{k}}{(z-w)^2}\d^{mn}
+\frac{1}{(z-w)}\{2C^{mn}_{~~k}\hat{A}^k -8\hat{G}^{mn}\} 
\en

We shall denote the  $\widehat{U(1)}$  current as 
$\hat{A}^0(z)$ 

\eq
\hat{A}^0(z)\hat{A}^0(w) = \frac{1/2}{(z-w)^2} + \hat{A}^0\hat{A}^0(w)
\en 

and the  8=1+7 free fermions  associated with the coset $SO(7)\times U(1)/G_2$ 
as $\psi^a(z)$ whose OPE's are the same as for the $N=8$ algebra.

Again one finds that there exists a unique solution to all the constraints of the
$N=7$ algebra which yields the following  expressions for its generators \cite{gk}

\eqn
G^A &=& -\frac{1}{4}(M^A)^{mn}\{ 
\hat{G}^{mn}+\bar{\psi}g^{mn}\psi \} 
\nn \\
S^m  &=&  \frac{i}{6}\sqrt{\frac{19}{6}} \{ C^m_{~np}\psi^n\hat{A}^p
 + 6i \psi^m\hat{A}^0 +3\pa\psi^m \} 
\\
T& =&  -\frac{1}{84}(\hat{G}^{mn} +\bar{\psi}g^{mn}\psi )^2  
- \frac{19}{7\cdot 12^2} \{ \hat{A}^m\hat{A}^m 
 -42 \hat{A}^0\hat{A}^0  \nn \\
&& +21i\pa\hat{A}^0 
+  18 \psi^m\pa\psi^m -16\hat{G}^{mn} (\bar{\psi}g^{mn}\psi) 
 +12i\hat{A}^m (\bar{\psi}\tilde{\g}^m\psi) \}
 \nn
\enn
where  $g^{mn} = \frac{1}{2} \g^{mn} + \frac{1}{8} C^{mn}_{~~pq} 
\g^{pq}$ and  $\psi^8=0$ . The $M^A$ are 
  the seven-dimensional representation matrices of $G_2$, $A=1,2,\ldots,14$,
  providing the explicit embedding of $G_2$ into 
$SO(7)$ \cite{gg,bo,gk}. They satisfy the identities \cite{bo,gk}

\eq
 tr (M^AM^B) = 2\d^{AB} 
\en
\[
M^A_{ij}M^A_{kl}= \frac{2}{3}(\d_{il}\d_{jk}-\d_{ik}\d_{jl})
-\frac{1}{3}C_{ijkl}
\]  \nn

It should be noted that the currents $\hat{A}^m$ are 
anti-hermitian which leads to factors of $i$ in the expressions for 
$S^m$ and $T$ so as to make all the terms appearing in them hermitian as 
required by unitarity. The above realization of the exceptional nonlinear 
$N=7$ algebra has central charge $c=89/12$ corresponding to $k=\hat{k}+1=3/2$.

\subsection{Realization in Terms of a Single Boson and $N$ Free Fermions}

In addition to the coset space solution given above to the most general Ansatz
for the realization of the exceptional nonlinear superconformal algebras there exist
another solution for each of these algebras in terms of a single boson and $N$ free
fermions.\cite{gk} For the $N=8$ algebra this realization leads to the following
expressions for the generators \cite{gk}

\eqn
 T^{mn} &=&  - \frac{i}{2} ( \bar{\psi}
 \tilde{\gamma}^{mn} \psi ) \nn \\ 
 S^m &=&  \frac{i}{\sqrt{5}} ( i\psi^m\hat{J}^8 +\frac{1}{3}
 C^{m}_{~npq}\psi^n\psi^p\psi^q 
+ C^{m}_{pq}\psi^p\psi^q\psi^8) 
\nn \\ 
 S^8 &=&  \frac{i}{\sqrt{5}}( -i\psi^8\hat{J}^8 
+ C_{mnp}\psi^m\psi^n\psi^p ) 
   \\
 T  & = &\frac{1}{20}\hat{J}^8\hat{J}^8-\frac{3}{8}\bar{\psi} \pa\psi 
+\frac{1}{240} (\bar{\psi}\tilde{\g}^{mn}\psi)(\bar{\psi}\tilde{\g}^{mn}\psi)
  \nn
\enn
with the central charge $c=26/5$ and level $k=1$ of the affine symmetry $\widehat{Spin}(7)$. 

For the $N=7$ algebra the corresponding realization gives

\eqn
G^A& = &-\frac{1}{3}(M^A)^{mn}( \bar{\psi}g^{mn}\psi) 
\nn \\
S^m & =& \frac{2}{\sqrt{6}}\{ 2\psi^m\hat{A}^0 
-\frac{i}{3}C^{m}_{npq}\psi^n\psi^p\psi^q \}  
\\
T &=& \frac{2}{3} \hat{A}^0\hat{A}^0 -\frac{2}{3} \psi^m\pa\psi^m  
+ \frac{1}{126} (\bar{\psi}g^{mn}\psi)(\bar{\psi}g^{mn}\psi) 
 \nn
\enn
In this case  the central charge is  $c=5$  with the level $k=1$ of the affine 
symmetry $\hat{G}_2$.

\section{An "Hybrid" Realization of the Exceptional Nonlinear Superconformal  
Algebras for Arbitrary Values of the Central Charge}
\setcounter{equation}{0}

The coset space method and the method of free fermions with a single boson
discussed above lead to realizations with a single definite value of the central charge 
in each case. Recently in a joint work with Behzad Bina \cite{bg}  we  have
developed a very general "hybrid" method for the unified realization of all
nonlinear superconformal algebras, quasi-superconformal algebras and $Z_2$
graded nonlinear superconformal algebras with both bosonic and fermionic
supersymmetry generators for arbitrary values of the central charge. In particular,
this method yields realizations of the exceptional nonlinear superconformal algebras
for all allowed  values of the central charge. 
The term "hybrid"  refers to the fact that in this method the susy generators involve 
free fermions in the coset of a supergroup , a scalar field ( dilaton) and  currents 
that form an affine Lie algebra isomorphic to the symmetry currents of the
nonlinear superconformal algebra. For this realization we have used the formulation
of nonlinear (quasi) superconformal algebras as given in \cite{fl}. In the formulation
of \cite{fl} the OPE of the supersymmetry generators take on the form

\eqn
G^{\alpha}(z) G^{\beta}(w)& =& \frac{b \d^{\alpha \beta}}{(z - w)^{3}} +
 \frac{\sigma \lambda^{\alpha \beta}_aJ^a  
(w)}{(z - w)^{2}} 
+ \frac{\sigma}{2} \frac{\lambda_a^{\alpha \beta}\partial 
J^{a}(w)}{(z - w)} \nn \\
&&+ \frac {2 \d ^{\alpha 
\beta}T(w)}{(z - w)}
 + \frac{\gamma P^{\alpha 
\beta}_{ab}J^{a}J^{b}(w)}{(z - w)} 
\enn

where 
$(\lambda_a)^{\alpha\beta} =\lambda_a^{\alpha\beta}  (  a,b,\ldots =1,2, \ldots , dim G)  $ are 
the  matrices of the representation under which the supersymmetry generators transform . The $J^a$ are the symmetry currents ( $Spin(7)$ and $G_2$ for the exceptional algebras). They satisfy 

\eq
J^{a}(z) G^{\alpha}(w)  = \frac{- \lambda^{a, \alpha}_{~~~\beta}
   G^{\beta}(w)}{(z-w)}  
\en 
\[
  J^{a}(z)J^{b}(w)  = -\frac{1}{2}\frac{k \ell ^{2} \eta ^{ab}}{(z-w)^{2}}
   +\frac{f^{ab}_{~~c}J^{c}(w)}{(z-w)} 
\] \nn

 where $\ell^{2} =4(6)$, and $\alpha,\beta, \ldots =1.2,\ldots, 8(7)$, for the $N=8(7)$ algebras ,respectively.
For the $N=8$ algebra one finds that the supersymmetry generators are of the form \cite{bg}

\eqn
  G_{\alpha}(z) &= &\frac{-2 i (k_0 +4)}{\sqrt{3k_0+15}}     \partial \psi _{\alpha}(z) + 
 \partial \phi    \psi_{\alpha} (z)
\nn \\
&& + \frac{i}{\sqrt{3k_0+15}} \{\lambda_{a,\alpha}^{~~~\beta} K^{a} \psi_{\beta}(z)
   +\frac{1}{3} \lambda_{a,\alpha}^{~~~\beta} \lambda^{a,\gamma \delta}
   \psi_{\gamma} \psi_{\delta} \psi_{\beta}(z) \}
\enn

where $K^a(z)$ are the currents of an affine $Spin(7)$ symmetry algebra of level $k_0$:

\eq
K^{a}(z)K^{b}(w) = \frac{-\frac{1}{2} k_{0} \ell^{2} 
   \eta^{ab}}{(z-w)^{2}} + 
   \frac{f^{ab}_{~~c}K^{c}(w)}{(z-w)}
\en

and $\psi^{\alpha}$  are eight free fermions

\eq
\psi ^{\alpha}(z) \psi^{\beta} (w) =  \frac{\d ^{\alpha 
   \beta}}{(z - w)} 
\en

 The scalar field $\phi (z)$ is normalized such that

\eq
\partial \phi (z) \partial \phi (w) = \frac{1}{(z - w)^{2}} + 
   \cdots .
\en 

The central charge of the $N=8$ superconformal algebra generated  by the above supersymmetry
generators is

$
c = \frac{(2k_0+1)^2}{(k_0+5)} + \frac{21k_0}{(k_0+5)} +5
$

which agrees with the general expression for the central charge given in section 2 by noting that the 'full' level $k$ of the 
affine $SO(7)$ symmetry is $k=k_0+1$.

For the $N= 7$ algebra the hybrid realization gives the following expression for the supersymmetry
generators \cite{bg}

\eqn
G_{\alpha}(z)& = &-\frac{i (3k_0+10)}{\sqrt{6(k_0+4)}} \partial \psi _{\alpha}(z) +  \partial \phi 
   \psi_{\alpha} (z)
\nn \\
&& + \frac{i}{\sqrt{6(k_0+4)}}
 \{ \lambda_{a,\alpha}^{~~~\beta} K^{a} \psi_{\beta}(z)
   +\frac{1}{3} \lambda_{a,\alpha}^{~~~\beta} \lambda^{a,\gamma \delta}
   \psi_{\gamma} \psi_{\delta} \psi_{\beta}(z) \}
\enn

where now the $K^a(z)$ generate an affine $G_2$ of level $k_0$ and we have seven free fermions
transforming in the $7$ of $G_2$. The central charge of the resulting $N=7$ algebra is

$
c = \frac{(3k_0+2)^2}{2(k_0+4)} + \frac{14k_0}{(k_0+4)} + \frac{9}{2}
$

 which again agrees with the general expression for the central charge given in section 2
by noting that $k=k_0+1$.

\end{document}